\documentclass[twocolumn,nobibnotes,aps,prb,showpacs,floats,reprint,10pt]{revtex4-1}
\usepackage{amssymb}

\usepackage{amsmath}
\usepackage[T1]{fontenc}
\usepackage{fourier}
\usepackage{graphicx}
\usepackage[colorlinks=true,
allcolors=blue,
dvipdfm=true]{hyperref}

\begin{document}

\title{Realization of the $\pi$-state in junctions formed by multi-band superconductors
with a spin-density-wave.}
\author{Andreas Moor, Anatoly F.~Volkov, and Konstantin B.~Efetov}
\affiliation{Theoretische Physik III,\\
Ruhr-Universit\"{a}t Bochum, D-44780 Bochum, Germany\\
}

\begin{abstract}
Using a simple model of multi-band superconductors, which can be applied in particular to Fe\nobreakdash-based pnictides, we calculate the Josephson
current~$I_{\text{J}}$ in a tunnel junction composed by such superconductors. We employ the tunneling Hamiltonian method and quasiclassical Green's functions. We study both the case of coexistence of the superconducting~($\Delta $) and magnetic~(SDW---spin density wave) order parameters and the case when only the superconducting order parameter exists. We show that the current~$I_{\text{J}}$ depends on the mutual orientation of magnetization of the SDW in the case of non-ideal nesting when the coexistence of superconducting and magnetic order parameters is possible as it takes place in Fe\nobreakdash-based pnictides. It is found that the realization of the $\pi$\nobreakdash-junction is possible in both cases. We compare our results for multi-band superconductors without the SDW with those obtained earlier and find that they coincide if the tunneling matrix elements are real. If these elements are complex, a new term appears in the formula for the Josephson critical current.
\end{abstract}

\date{\today}
\maketitle

\textit{Introduction.} The superconducting order parameter~(OP) in
the BCS model~$\Delta$ is either constant or angle-dependent as it occurs in anisotropic superconductors.~\cite{Tinkham} Superconductors of new types discovered in last decades are characterized by a more complicated~OP. For example, the OP in high\nobreakdash-$T_{\text{c}}$ superconductors depends on
angles and in some of them corresponds to the singlet, so-called $d$\nobreakdash-wave pairing. This means
that the OP changes sign in certain directions and therefore nodes arise in the excitation spectra~\cite{HighT_c}.

The OP in the recently discovered superconducting materials---Fe\nobreakdash-based
pnictides, which have rather high~$T_{\text{c}}$~($\sim 60K$)~\cite{Pnictides}---is
also nontrivial. These materials belong to a class of multi-band
superconductors; their band structure consists of electron and hole bands.
The OP in these bands may have not only different amplitudes but also
opposite signs. If the OP in the hole and electron bands have different
signs, one speaks of the $s_{+-}$\nobreakdash-pairing---in contrast to the the
$s_{++}$\nobreakdash-pairing in case of the same signs of the OP in different bands (see, for
example, reviews~\onlinecite{Review1,Review2,Review3,Review4,Review5,Review5a,Review6}).
Investigating the structure of the OP in different types of superconductors
is a very important task since this study may shed light on the mechanism of
superconductivity in these materials.

One of the effective methods to determine the structure of the OP is the
measurement of the Josephson current. For example, the sign change of the
OP in high\nobreakdash-$T_{\text{c}}$ superconductors has been proven in experiments, in which
the critical Josephson current~$I_{\text{c}}$ was measured in a setup containing
two Josephson junctions connected by a superconducting loop (i.e., on SQUID)~\cite{HighT_c}.
The dc Josephson effect in multi-band
superconductors has been studied theoretically in many papers~\cite{Mazin95,Agterberg,Brinkman,Nagaosa,Chen,Linder,Ota,Zhang,Lin}.
The calculations in Refs.~\onlinecite{Agterberg,Brinkman} are focused on the multi-band superconductor $\text{MgB}_{2}$,
whereas the main attention of the authors of Refs.~\onlinecite{Nagaosa,Chen,Linder,Ota,Zhang,Lin} is paid to the
Fe\nobreakdash-based pnictides. Agterberg et al.~\cite{Agterberg} have shown that the Josephson
$\text{S/I/S}_{\text{mb}}$ junctions may have a
negative critical current~$I_{\text{c}}$ if the OP~$\Delta$ is negative in some
of the bands (here,~$\text{S}$ and~$\text{S}_{\text{mb}}$ mean single-band and multi-band
superconductors, respectively, $\text{I}$~stands for an insulating layer). This idea allows a
simple physical interpretation. The Josephson current~$I_{\text{c}}$ in an
$\text{S/I/S}_{\text{mb}}$~junction can be written as ${I_{\text{c}} \propto \sum_{\alpha} \Delta \Delta_{\alpha } / R_{\alpha}}$,
where $R_{\alpha}$~is the resistance for electron transitions from the $\text{S}$~superconductor to
the band~$\alpha$ in the $\text{S}_{\text{mb}}$~superconductor. It is clear that if $\Delta_{\alpha}$ is negative, for
instance in the band with~${\alpha = 1}$, the current~$I_{\text{c}}$ may be also
negative. This happens provided the resistance~$R_{1}$ is sufficiently small. The
ground state of the Josephson junction with negative~$I_{\text{c}}$ is called the~$\pi$\nobreakdash-state.

Note that the existence of the $\pi$\nobreakdash-state of the Josephson junction is
interesting in itself because such junctions can be used in practical
applications (see, e.g., Ref.~\onlinecite{Mints} and references therein). The $\pi$\nobreakdash-state
is realized in~$\text{S/F/S}$ Josephson junctions and is being studied very
actively (for reviews see Refs.~\onlinecite{GolubovRMP,BuzdinRMP,BVErmp,EschrigPhToday}). Therefore, there is
a need both from the point of view of fundamental research and of the future
applications to study the possible realizations of the $\pi$\nobreakdash-state
in Fe\nobreakdash-based pnictides. In
all publications mentioned above the presence of the spin density wave~(SDW)
in these materials is ignored. On the other hand, it is known that there is
a region on the $T$\nobreakdash-$x$\nobreakdash-plane (temperature and doping level)
where the superconducting and magnetic~(SDW) phases coexist.

In the present paper we calculate the Josephson current~$I_{\text{c}}$ in tunnel
junctions formed by multi-band superconductors with and without the SDW, i.e., in the
${\text{(S}_{\text{mb}} + \text{SDW)}\text{/I/(S}_{\text{mb}}+\text{SDW)}}$ or ${\text{S}_{\text{mb}}\text{/I/S}_{\text{mb}}}$ junctions, where ${\text{(S}_{\text{mb}} + \text{SDW)}}$ stands for a multi-band superconductor with a spin density wave. We show that in junctions with the SDW the critical current~$I_{\text{c}}$ consists of
two terms. The first one is proportional to $\Delta_{\text{l}} \Delta_{\text{r}}$ and
does not depend on the angle between the magnetization vectors
$\mathbf{m}_{\text{l},\text{r}}$ in the SDW on the left~(l) and on the right~(r).
The second term is proportional to $\Delta_{\text{l}} \Delta_{\text{r}} (\mathbf{m}_{\text{l}} \cdot \mathbf{m}_{\text{r}})$.
This means that this component can be negative and, as we will show,
it can prevail so that ${I_{\text{c}} \propto \cos(2 \theta)}$, where~$2 \theta$ is
the angle between the vectors~$\mathbf{m}_{\text{l}}$ and~$\mathbf{m}_{\text{r}}$.
Thus,~$I_{\text{c}}$ may be negative in such junctions. We also find the Josephson current in the $\text{S}_{++}\text{/I/S}_{++}$, $\text{S}_{+-}\text{/I/S}_{+-}$ and $\text{S}_{++}\text{/I/S}_{+-}$ junctions in the absence of the SDW.

\textit{System under consideration. Model.} We consider a tunnel
$\text{S}_{\text{mb}}\text{/I/S}_{\text{{mb}}}$ junction. Each superconductor on the
left and on the right is described by the Hamiltonian~$\mathcal{H}_{\text{l},\text{r}}$ which contains
the superconducting and magnetic energies taken in the mean field
approximation~\cite{Chubukov09,Schmalian10,Moor11}. Transitions
of electrons between superconductors is described by the tunneling
Hamiltonian
\begin{equation}
\mathcal{H}_{\text{T}} = \sum_{\mathbf{p}, \alpha, \beta} \{ \mathcal{T}_{\alpha \beta}
\hat{a}_{\alpha,\text{r}}^{+} \hat{a}_{\beta,\text{l}} + \text{H.c.} \} =
\sum_{\mathbf{p}} \{ \hat{C}_{\text{r}}^{\dagger} \hat{\mathrm{H}}_{\text{T}} \hat{C}_{\text{l}} + \text{H.c.} \} \,,
\label{TunHam}
\end{equation}%
where the matrix elements~$\mathcal{T}_{\alpha \beta }$ describe the
electron tunneling between the same bands if ${\alpha = \beta}$,
${\mathcal{T}_{\alpha \alpha} \equiv \mathcal{T}_{\alpha}}$, and between different bands
if ${\alpha \neq \beta}$. In the latter case one has ${\mathcal{T}_{\alpha
\beta} = \mathcal{T}_{\beta \alpha }^{\ast}}$ for the identical
superconductors at the left and right. We assume that the matrix elements~$\mathcal{T}_{\alpha \beta}$
do not depend on momentum~$\mathbf{p}$ counted from the center of the valley. The band~${\alpha = 1}$ (resp.~$2$) is assumed to be the hole (resp.~electron) band. The ${\hat{C} \equiv \hat{C}_{\alpha ns}}$ operators are related to the
$\hat{a}_{\alpha}$~operators as follows: ${\hat{C}_{\alpha ns} = A_{ns}}$ for ${\alpha =1}$ and
${\hat{C}_{\alpha ns} = B_{ns}}$ for ${\alpha = 2}$. In the hole band one has
${\hat{A}_{1n \uparrow} = \hat{a}_{1\downarrow}^{\dagger}}$ for ${n=1}$ and
${\hat{A}_{1n\uparrow} = \hat{a}_{1\uparrow}}$ for ${n=2}$. In the electron band
the relations ${\hat{B}_{2n\uparrow} = \hat{a}_{2\uparrow}}$ for ${n=1}$ and
${\hat{B}_{2n\uparrow} = \hat{a}_{2\downarrow}^{\dagger}}$ for ${n=2}$ take place.
One can see that the labels~$\alpha$, $n$~and~$s$ are the band, Gor'kov-Nambu and spin
indices, respectively.

The operator $\hat{\mathrm{H}}_{\text{T}}$ can be written in terms of the matrices
$\hat{\rho}$, $\hat{\tau}$~and~$\hat{\sigma}$ operating in the band,
Gor'kov-Nambu and spin spaces: ${\hat{\mathrm{H}}_{\text{T}} = \hat{\Gamma} \cdot \hat{\tau}_{3}}$.
Here, the matrix~$\hat{\Gamma}$ is given by
\begin{equation}
\label{Gamma}
\hat{\Gamma} = \frac{1}{2} \left[ \mathcal{T}_{+} \hat{X}_{300} + \mathcal{T}_{-}
- \mathrm{i} \left( \mathcal{V}^{\prime} \hat{X}_{210} -
\mathcal{V}^{\prime \prime} \hat{X}_{220} \right)
\right] \,,
\end{equation}
where ${\mathcal{T}_{\pm} = (\mathcal{T}_{1} \pm \mathcal{T}_{2})/2}$ and
${\mathcal{V} \equiv \mathcal{V}^{\prime} + \mathrm{i} \mathcal{V}^{\prime \prime} = \mathcal{T}_{12}}$.
The matrices~$\hat{X}_{\alpha n s}$ are defined as ${\hat{X}_{\alpha n s} = \hat{\rho}_{\alpha} \cdot \hat{\tau}_{n} \cdot \hat{\sigma}_{s}}$.
It is worth making an important note. As is
known, in tunnel junctions composed by single band superconductors, the
relation ${\mathcal{T}(p,p^{\prime}) = \mathcal{T}^{\ast}(-p,-p^{\prime})}$ holds which
is a consequence of the time reversal symmetry. Therefore, the matrix
elements~$\mathcal{T}_{1,2}$ that describe tunneling between identical bands
are real if we assume that these matrix elements do not depend on momenta~$p$ and~$p^{\prime}$.
However, the matrix element~${\mathcal{V} \equiv \mathcal{T}_{12}}$ describes tunneling between different bands. In
this case, the initial and final states are different and the idea about time
reversal symmetry is not applicable. Moreover, the matrix elements~$\mathcal{T}_{\alpha \beta}$ are determined by the nature of scattering at the interface, which can be time-reversal symmetry breaking. Within the limits of the approach used we only can treat the matrix elements~$\mathcal{T}_{12}$ as phenomenological input parameters and consider them as complex in order to keep the description as general as possible, i.e.,~${\mathcal{V} = \Re(\mathcal{V}) + \mathrm{i} \Im(\mathcal{V})}$.

One can obtain the Eilenberger-like equation
for quasiclassical Green's functions in the, e.g., right electrode
with the self-energy part~$\hat{\Sigma}_{\text{T},\text{r}}$ which is related to the
tunneling Hamiltonian:
${\hat{\Sigma}_{\text{T},\text{r}} = \hat{\Gamma} \cdot \hat{g}_{\text{l}} \cdot \hat{\Gamma}^{\dagger}}$~\cite{Moor12}. The matrix $\hat{g}_{\text{l}}$ in the self-energy part~$\hat{\Sigma}_{\text{T},\text{r}}$ is the quasiclassical
Green's function in the left electrode. It can be found from the
Eilenberger equation neglecting the self-energy~$\hat{\Sigma}_{\text{T},\text{r}}$ since we assume
small tunneling probability (the method of
quasiclassical Green's functions is described in
reviews~\onlinecite{Rammer,BelzigRev,Kopnin,Kamenev}). In the case of ideal nesting,
this matrix function is given in Ref.~\onlinecite{Moor12}. If  the nesting is not
ideal, this function acquires a more complicated form.

\textit{Josephson current.} From the generalized Eilenberger
equation one obtains the rate of the charge variation with time, e.g., in the right electrode:
${(\partial_t Q_{\text{r}} + \partial_{t^{\prime}}
Q_{\text{r}})|_{t = t^{\prime}} = I_{\text{T}}}$, where ${Q_{\text{r}} = e N_{\text{r}}(0)
\int \mathrm{d} \epsilon \mathrm{Tr} \{\hat{X}_{300} \hat{g}_{\text{r}}^{K} \}}$ with the density of
states at the Fermi level~$N_{\text{r}}(0)$ and~$\hat{g}_{\text{r}}^{K}$ is the Keldysh component of the Green's function. In
equilibrium, this function is equal to
${\hat{g}_{\text{r}}^{K} = (\hat{g}_{\text{r}}^{R} - \hat{g}_{\text{r}}^{A}) \tanh (\epsilon \beta )}$
with $\beta = (2T)^{-1}$. The tunneling current in equilibrium is the nondissipative
Josephson current~$I_{\text{J}}$. It is given by
\begin{equation}
I_{\text{J}} = c_{1} (4 \pi \mathrm{i} T) \sum_{\omega} \mathrm{Tr} \langle \hat{X}_{330} [\hat{\Gamma} \hat{g}_{\text{l}}(\omega) \hat{\Gamma} \,, \hat{g}_{\text{r}}(\omega)] \rangle \,,  \label{JosCurrent}
\end{equation}
where $c_{1} = \pi e N_{\text{l}}(0) N_{\text{r}}(0) / 16$, the angle brackets~${\langle \ldots \rangle}$ mean the angle averaging in the momentum space and $\hat{g}_{\text{l},\text{r}}(\omega )$ are the
Green's functions in the left (right) electrodes in the Matsubara representation. This expression for the Josephson current is rather general. It is applicable both to systems with an SDW (e.g.,~Fe\nobreakdash-based pnictides with non-ideal nesting) and to systems, where only the superconducting order parameter exists in different bands. The quasiclassical Green's functions~$\hat{g}$ have different form in these systems.

\textit{SC without SDW.} First, using Eqs.~(\ref{Gamma}--\ref{JosCurrent}) we
calculate the Josephson current in a tunnel junction formed by multi-band superconductors with no magnetic order. The quasiclassical Green's function~${\hat{g} = g \hat{X}_{030} + \hat{f}}$ consists of the normal (the first term on the right) and the condensate (Gor'kov's) function~$\hat{f}$. In the absence of a voltage, the trace in Eq.~(\ref{JosCurrent})
is not zero only for the condensate component, which has here the
form\cite{Moor11} (we assume ${|\Delta_{\text{hole}}| = |\Delta_{\text{electron}}| \equiv \Delta}$)
\begin{equation}
\hat{f}_{\text{l},\text{r}}(\omega) = \frac{\Delta}{\mathcal{D}}
\begin{cases}
\hat{X}_{323} \cos (\varphi/2) \pm \hat{X}_{013} \sin (\varphi/2) & \text{for $s_{+-}$\nobreakdash-pairing} \,, \\
\hat{X}_{023} \cos (\varphi/2) \pm \hat{X}_{313} \sin (\varphi/2) & \text{for $s_{++}$\nobreakdash-pairing} \,,
\end{cases}
\end{equation}
where ${\mathcal{D} = \sqrt{\omega^{2} + \Delta^{2}}}$ and~$\varphi$ is the phase difference between the left and the right superconductors, set to be the phase difference between the order parameters in the electron bands; then, the phase of the order parameter in the hole band is~${\varphi_{\text{h}} = \pi - \varphi/2}$. We calculate the current~$I_{\text{J}}$ for different junctions consisting of the same materials of the left and right electrodes: a)~$\text{S}_{++}\text{/I/S}_{++}$, b)~$\text{S}_{+-}\text{/I/S}_{+-}$ and c)~$\text{S}_{++}\text{/I/S}_{+-}$. The only difference between the superconductors~$\text{S}_{++}$ and~$\text{S}_{+-}$ is that the phases in the bands~1 and~2 in~$\text{S}_{+-}$ differ by~$\pi$, meaning that the order parameters~$\Delta_1$ and~$\Delta_2$ have opposite signs, ${\Delta_1 = - \Delta_2 \equiv \Delta}$. In the symmetric cases of identical superconductors forming the junction we obtain the standard formula $I_{\text{J}} = I_{\text{c}} \sin \varphi$ with different critical current
\begin{equation}
I_{\text{c}}/I_{0} = |\mathcal{T}|^{-2}
\begin{cases}
|\mathcal{T}_{1}|^{2} + |\mathcal{T}_{2}|^{2} + 2 \Re ( \mathcal{V}^{2} ) & \text{for S}_{++}\text{/I/S}_{++} \,, \\
|\mathcal{T}_{1}|^{2} + |\mathcal{T}_{2}|^{2} - 2 \Re ( \mathcal{V}^{2} ) & \text{for S}_{+-}\text{/I/S}_{+-}  \,,
\end{cases}
\label{CritCurrent}
\end{equation}%
where ${I_{0} = (\pi \Delta /2 e R_{\text{n}}) \tanh (\Delta /2T)}$, ${|\mathcal{T}|^{2} = |\mathcal{T}|_{1}^{2} + |\mathcal{T}|_{2}^{2} + |\mathcal{V}|^{2}}$ and
${R_{\text{n}}^{-1} = 4 \pi e^{2} N_{\text{l}}(0) N_{\text{r}}(0) \left(|\mathcal{T}|_{1}^{2} + |\mathcal{T}|_{2}^{2} + 2 |\mathcal{V}|^{2}\right)}$
is the resistance of the junction in the normal state.
In the case of an asymmetrical $\text{S}_{++}\text{/I/S}_{+-}$ junction
we obtain a quite different result
\begin{equation}
I_{\text{J}} / I_{0} =  |\mathcal{T}|^{-2} \left[ \left( |\mathcal{T}_{1}|^{2} - |\mathcal{T}_{2}|^{2} \right) \sin \varphi + \Im (\mathcal{V}^{2}) \cos \varphi \right]  \,.
\label{JosCur++/+-}
\end{equation}

As follows from Eqs.~(\ref{CritCurrent}) and~(\ref{JosCur++/+-}), in case of real tunneling matrix
elements~$\mathcal{T}_{1,2}$ and~$\mathcal{V}$, the critical current
may be negative in junctions~$\text{S}_{+-}\text{/I/S}_{+-}$ and~$\text{S}_{++}\text{/I/S}_{+-}$. In
the first case, $I_{\text{c}}$~is negative if the interband transitions dominate,
i.e.,~$\mathcal{T}_{1,2} \ll \mathcal{V}$. In the S$_{++}$/I/S$_{+-}$
junction, $I_{\text{c}}$~changes sign at ${ |\mathcal{T}_{1}|^{2} < |\mathcal{T}_{2}|^{2}}$.
These results resemble those obtained in
Refs.~\onlinecite{Agterberg,Brinkman,Chen,Linder,Ota,Zhang,Lin}, where the Josephson current in
junctions of the type $\text{S}\text{/I/S}_{\text{mb}}$ has been studied.

However, if the tunneling matrix elements~$\mathcal{T}_{12}$ are complex, the obtained results
do not reduce to those established earlier~\cite{Agterberg,Brinkman,Nagaosa,Chen,Linder,Ota,Zhang,Lin}. Especially
interesting is the result for the critical current in an $\text{S}_{++}\text{/I/S}_{+-}$
junction. In this case, the Josephson current has the form:
${I_{\text{c}} \propto I_{\text{c} 1} \sin (\varphi ) + I_{\text{c} 2} \cos (\varphi)}$.
This means that a spontaneous current arises even at zero phase difference; or a finite phase difference is established across the junction in a disconnected circuit. The presence of a spontaneous condensate current
has been established earlier in different systems where the time-reversal
symmetry breaking~(TRSB) takes place. For example, this current arises in a
superconducting loop containing a $\pi$\nobreakdash-Josephson
junction~\cite{Bulaev77,Ryazanov07} or in SF bilayer~\cite{Annett02}. Note that there is a similarity of
this effect and the anomalous proximity effect in $\text{S}_{+-}$ system studied in Refs.~\onlinecite{Nagaosa,Lin,Koshelev1}.\footnote{Note that Koshelev\cite{Koshelev2} considered another mechanism of the TRSB taking into account tunneling processes of higher order. Our calculations are restricted
by the lowest order processes.} In these papers, tunnel
junctions of the $\text{S}_{+-}\text{/I/S}$ type were considered, and it was
shown that at some values of coupling constant between different bands, a
finite phase difference~$\varphi$ not equal to~$\pi$ arises in the system in the ground state in the absence of the total current. The mechanism of the appearance of a finite phase difference~$\varphi_{0}$ in our system is completely different. First, we consider an~$\text{S}_{++}\text{/I/S}_{+-}$ junction with multi-band superconductors on both sides. Second, in our case, the phase difference~$\varphi_0$ arises only if the interband transitions take place and the probability amplitude~$\mathcal{V}$ for these transitions has a non-vanishing imaginary part, i.e.,~${\Im(\mathcal{V}) \neq 0}$.
In the frames of the tunneling Hamiltonian method~(THM) one can not estimate the magnitude of ${\mathcal{V}}$ because the tunneling matrix elements in this theory are considered as phenomenological parameters (in order to calculate them, one has to go beyond the THM). On the other hand, there are no reasons to regard them as real quantities.

\textit{SC with SDW.} Now, we consider the most interesting
case of the non-ideal nesting when the superconducting~($\Delta $) and
magnetic~(SDW) order parameter may coexist in a certain interval of
doping level~$x$ and temperature~$T$~\cite{Chubukov09,Schmalian10},
i.e., we introduce a parameter $\delta \mu _{\mathbf{p}}$, describing the mismatch of the effective Fermi surfaces of the bands. Then, the Green's function for the $s_{+-}$\nobreakdash-pairing, in case ${\varphi = 0}$
and the magnetization vector oriented along the $z$\nobreakdash-axis, has the form
$\hat{g}_{+-} \equiv \hat{g}_{+-}(0,0) = g_{030} \hat{X}_{030} + g_{100} \hat{X}_{100} + g_{123} \hat{X}_{123} + g_{213} \hat{X}_{213} + g_{300} \hat{X}_{300} + g_{323} \hat{X}_{323}$. For the $s_{++}$\nobreakdash-pairing
we have $\hat{g}_{++} \equiv \hat{g}_{++}(0,0) = \tilde{g}_{023} \hat{X}_{023} + \tilde{g}_{030} \hat{X}_{030} + \tilde{g}_{123} \hat{X}_{123} + \tilde{g}_{130} \hat{X}_{130} + \tilde{g}_{213} \hat{X}_{213} + \tilde{g}_{300} \hat{X}_{300}$. That is, the matrices~$\hat{g}_{+-}$ and~$\hat{g}_{++}$ have a rather complicated
form. In our notations for the $\text{S}_{+-}$~system we follow
previous papers~\cite{Moor11, Moor12} and the quantities with tilde denote
the according ones in the $\text{S}_{++}$~system. However, only two
terms---$g_{323}\hat{X}_{323}$, $g_{100}\hat{X}_{100}$ in $\hat{g}_{+-}$ and $g_{023}\hat{X}_{\text{0}23}$, $g_{130}\hat{X}_{130}$ in $\hat{g}_{++}$---contribute to the Josephson current.
\begin{figure}
\begin{center}
\includegraphics[width=0.23\textwidth]{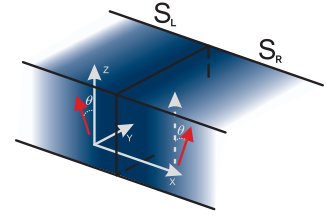}
\end{center}
\caption{(Color online.) Considered setup of the Josephson junction.}
\label{fig:Setup}
\end{figure}

If the phases~$\varphi/2$ are different~($\pm \varphi/2$) and the angle~${\theta \neq 0}$ (see Fig.~\ref{fig:Setup}), the Green's
functions in the left (right) superconductors~$\hat{g}_{\text{l},\text{r}}(\varphi, \theta)$
are expressed through the matrices~$\hat{g}(0,0)$ with the help of the
unitary transformations: ${\hat{g}_{\text{l},\text{r}}(\varphi, \theta) = \hat{R}_{\pm \theta
} \cdot \hat{S}_{\pm \varphi} \cdot \hat{g}(0,0) \cdot \hat{R}_{\pm \theta}^{\dagger} \cdot
\hat{S}_{\pm \varphi}^{\dagger}}$, where the signs~$\pm$ relate to the left~(right)
electrodes. The transformation matrices are: ${\hat{S}_{\pm \varphi} = \exp (\pm \mathrm{i} \hat{X}_{330} \varphi / 4)}$ and ${\hat{R}_{\pm \theta} = \exp (\pm \mathrm{i} \hat{X}_{331} \theta / 2)}$. They can be called the rotation matrices
in the Gor'kov--Nambu and spin spaces.

In the symmetric cases of $\text{S}_{++}\text{/I/S}_{++}$ and~$\text{S}_{+-}\text{/I/S}_{+-}$ junctions
we obtain for the Josephson current~${I_{\text{J}} = I_{\text{c}} \sin \varphi}$ with
\begin{align}
\label{CritCurPlusPlus}
I_{\text{c}} / I_{0} \propto |\mathcal{T}|^{-2} \sum_{\omega} & \left\langle \tilde{g}_{023}^{\prime 2} \left[ |\mathcal{T}_{1}|^{2} + |\mathcal{T}_{2}|^{2} + 2 \Re(\mathcal{V}^{2}) \right] \right. \\
&\left. + \tilde{g}_{130}^{\prime 2} \cos (2 \theta ) (\mathcal{T}_{1} \mathcal{T}_{2} - |\mathcal{V}|^{2}) \right\rangle \notag
\end{align}
and
\begin{align}
\label{CritCurPlusMinus}
I_{\text{c}} / I_{0} \propto |\mathcal{T}|^{-2} \sum_{\omega} & \left\langle g_{323}^{\prime 2} \left[ |\mathcal{T}_{1}|^{2} + |\mathcal{T}_{2}|^{2} - 2 \Re(\mathcal{V}^{2}) \right] \right. \\
&\left. + g_{100}^{\prime 2} \cos (2 \theta ) (\mathcal{T}_{1} \mathcal{T}_{2} - |\mathcal{V}|^{2}) \right\rangle \,, \notag
\end{align}
respectively, where $\tilde{g}_{023}^{\prime} = \tilde{\Delta} \zeta_{+}^{-1} |\psi|^{-2} \Im \big\{ \big( \tilde{W}_{M0}^2 - \delta \tilde{\mu}_{\mathbf{p}}^2 + \zeta_{+}^2 \big) \psi \big\} $, $\tilde{g}_{130}^{\prime} = \omega_{n} \tilde{\Delta} \tilde{W}_{M0} \zeta_{+}^{-1} |\psi|^{-2} \Im\{ \psi \} $, $g_{100}^{\prime} = \Delta W_{M0}\zeta ^{-1}|\chi|^{-2} \Im \{\chi\}$ and $g_{323}^{\prime} = \Delta \zeta ^{-1}|\chi|^{-2} \big( \zeta \Re \{\chi\} + \delta \mu_{\mathbf{p}} \Im \{ \chi \} \big)$, with $\psi = \sqrt{\tilde{W}_{M0}^2 + \tilde{\Delta}^2 + \omega_{n}^2 - \delta \tilde{\mu}_{\mathbf{p}}^2 - 2 \zeta_{+}}$, $\zeta_{+} = -\mathrm{i} \sqrt{(\delta \tilde{\mu}_{\mathbf{p}}^2 - \tilde{W}_{M0}^2) \tilde{\Delta}^2 + \omega^2 \delta \tilde{\mu}_{\mathbf{p}}^2}$, $\chi = \sqrt{W_{M0}^{2} + \big( \zeta + \mathrm{i}\delta \mu_{\mathbf{p}} \big)^{2}}$ and $\zeta = \sqrt{\Delta^{2} + \omega_{n}^{2}}$. We see that the angle-independent part in Eqs.~(\ref{CritCurPlusPlus}) and~(\ref{CritCurPlusMinus}) (the terms in the square brackets) qualitatively has the same form as in Eq.~(\ref{CritCurrent}), but the numerical factor~($\tilde{g}_{023}^{\prime}$) is different and depends on the deviation from the ideal nesting~$\delta \mu_{\mathbf{p}}$.

However, in the case of the non-ideal nesting,~${\delta \mu_{\mathbf{p}} \neq 0}$, a new term proportional to $\cos (2\theta )$ arises in the expression for the critical Josephson current. This means that the Josephson current depends on mutual orientation of the magnetization vectors in the SDW. This term in the current survives even if all the tunnel matrix elements are real.

In the case of the~$\text{S}_{++}\text{/I/S}_{+-}$ junction we obtain no angle dependent part, and the formula for~$I_{\text{J}}$ looks similar to the corresponding expression for the case of
the absence of the SDW order but with modified coefficients due to the presence of the SDW and the non-ideal nesting, i.e.,
\begin{align}
I_{\text{J}} / I_{0} \propto & |\mathcal{T}|^{-2} \sum_{\omega} \langle g_{013}^{\prime} \tilde{g}_{023}^{\prime} \\
&\times \left( \left[ |\mathcal{T}_{1}|^{2} - |\mathcal{T}_{2}|^{2} \right] \sin \varphi + \Im ( \mathcal{V}^{2} ) \cos \varphi  \right)\rangle \,. \notag
\label{CritCurMixed}
\end{align}

\textit{$\pi$\nobreakdash-state.} In principle, the realization of the $\pi$\nobreakdash-state is possible in all three cases. The case~c) of the $\text{S}_{++}\text{/I/S}_{+-}$ junction is discussed above. In the cases~a) and b), i.e., in the $\text{S}_{++}\text{/I/S}_{++}$ and $\text{S}_{+-}\text{/I/S}_{+-}$ junctions it is possible to have the change of the sign of the critical current near~$T_{\text{c}}$ for sufficiently large ratio of~$T_{\text{s}}/T_{\text{c}}$, where $T_\text{s}$ is the transition temperature for the SDW to normal state.

However, the $\pi$\nobreakdash-state is most easily realized and observed in the $\text{S}_{+-}\text{/I/S}_{+-}$ junction,
where the~$\cos (2 \theta)$ term plays an important role. This, together with the mechanically favorable properties of the Fe\nobreakdash-pnictides~\cite{Togano,Katase}, makes the $\text{S}_{+-}\text{/I/S}_{+-}$ junction very attractive for possible applications in the quantum devices. In particular, the possibility of creating wires allows one to think of the realization of the so-called $\phi$\nobreakdash-junction (arbitrary phase shift, not only~$0$ or~$\pi$) out of pieces of Fe\nobreakdash-pnictide wire, put together mutually rotated and separated by an insulating layer---a chain of Josephson junctions.

\textit{Discussion.} Starting from a simple model of multiband superconductors with or without an SDW (this model is applicable to the
Fe\nobreakdash-based pnictides) and a tunneling Hamiltonian, we calculated the dc Josephson current in the junctions consisting of such superconductors separated by an insulating layer: a)~$\text{S}_{++}\text{/I/S}_{++}$, b)~$\text{S}_{+-}\text{/I/S}_{+-}$ and c)~$\text{S}_{++}\text{/I/S}_{+-}$, where the indices indicate the pairing of the superconducting OP of the system. We focused mainly on the case when the superconducting OP coexists with the SDW, but for completeness we analyzed the case of the absence of the SDW. Even in this case the Josephson current has an unusual phase dependence if the interband tunneling matrix element $\mathcal{V}$ has a non-vanishing imaginary part.

The most interesting result is obtained for the case of coexistence of the SC and SDW order parameters (non-ideal nesting in Fe\nobreakdash-based pnictides). In this case the critical current~$I_{\text{c}}$ contains a term~$I_{\text{c}}(\theta)$ that depends on the angle between magnetization vectors in the SDW at the left and right side---${I_{\text{c}}(\theta) \propto \cos (2 \theta)}$. This term allows the identification of the $s_{+-}$ pairing and, on the other hand, the realization of the $\pi$\nobreakdash-state of the $\text{S}_{+-}\text{/I/S}_{+-}$ junction.

\bigskip

\textit{Acknowledgments.} The authors are grateful to I.~Eremin, F.~Nogueira and S.~Syzranov for useful remarks and discussions.

\end{document}